\documentclass{osa-article}

\usepackage{amsfonts}
\usepackage{amsmath}
\usepackage{amssymb}
\usepackage{booktabs}
\usepackage{tabularx}
\usepackage{mathtools}
\usepackage{pgfplots}
\usepackage{geometry}
\usepackage{pdflscape}
\usepackage{enumerate}
\usepackage{esvect}
\usepackage[figuresright]{rotating}
\usepackage{footnote}
\usepackage{listings}
\usepackage{xcolor}
\usepackage{enumitem}
\usepackage{tocloft}
\usepackage{textcomp}
\usepackage{braket}
\usepackage{nicefrac}
\usepackage{nccmath}
\usepackage{bm}
\usepackage{import}
\usepackage{dpfloat}
\usepackage{nicefrac}
\usepackage{subcaption}
\usepackage{float}
\usepackage{afterpage}
\usepackage{multirow}
\usepackage{siunitx}
\usepackage{graphicx}
\usepackage{tikz}
\usepackage[T1]{fontenc}
\usepackage{xr-hyper}
\usepackage{changes}
\usepackage{hyperref}
\DeclarePairedDelimiter\abs{\lvert}{\rvert}%

\makesavenoteenv{tabular}

\graphicspath{{Figs/}}

\journal{osajournal}

\articletype{Research Article}

\begin{document}
    
    \title{Improving Holographic Search Algorithms using Sorted Pixel Selection}
    
    \author{Peter J. Christopher, \authormark{1,*} Jamie D. Lake, \authormark{2} Daoming Dong,\authormark{1} Hannah J. Joyce\authormark{2} and Timothy D. Wilkinson\authormark{1}}
    
    \address{\authormark{1}Centre of Molecular Materials, Photonics and Electronics, University of Cambridge\\
    \authormark{2}Electronic + Photonic Nanomaterials, University of Cambridge}

    \email{\authormark{*}pjc209@cam.ac.uk}
    
    \homepage{http:\textbackslash\textbackslash www.peterjchristopher.me.uk}

    \begin{abstract}
        Traditional search algorithms for computer hologram generation such as Direct Search and Simulated Annealing offer some of the best hologram qualities at convergence when compared to rival approaches. Their slow generation times and high processing power requirements mean, however, that they see little use in performance critical applications. 
        
        This paper presents the novel Sorted Pixel Selection~(SPS) modification for Holographic Search Algorithms~(HSAs) that offers Mean Square Error~(MSE) reductions in the range of $14.7-19.2\%$ for the test images used. SPS operates by substituting a weighted search selection procedure for traditional random pixel selection processes. While small, the improvements seen are observed consistently across a wide range of test cases and require limited overhead for implementation.  
    \end{abstract}

    \section{Introduction}
    
    Computer Generated Holography~(CGH) is a widely used component of technologies for imaging~\cite{Daneshpanah2010,Sheen2001}, displays~\cite{Wu1996,Kuo1997,Maimone2017,Yamada2018}, lithography~\cite{Turberfield2000}, beam shaping, optical tweezing~\cite{Melville2003,Grieve2009} and telecommunications~\cite{HoloAppTeleCom}. For applications where hologram quality rather than generation speed is fundamental, holographic search algorithms~(HSAs) are the primary approach used for hologram generation~\cite{DirectSearch_2}.
    
    This paper presents a novel means for improving the performance of HSAs, Sorted Pixel Selection~(SPS). While the approach is general in scope, the two most popular HSAs are chosen as test cases: Direct Search~(DS) and Simulated Annealing~(SA). 
    
    A very brief background to CGH is presented, and the details of DS and SA are explained. Some heuristic observations of the properties of the process are used to justify the explanation for the SPS approach before the technique itself is presented and explained.
        
    \section{Background} \label{background}
    
    A two-dimensional Discrete Fourier Transform~(DFT) is given by
    
    \begin{align}
        F_{u,v} = \mathcal{F}\{f_{x,y}\}         & = \frac{1}{\sqrt{N_xN_y}}\sum_{x=0}^{N_x-1}\sum_{y=0}^{N_y-1} f_{xy}e^{-2\pi i \left(\frac{u x}{N_x} + \frac{v y}{N_y}\right)} \label{fouriertrans2d5c}   \\
        f_{x,y} = \mathcal{F}^{ - 1 }\{F_{u,v}\} & = \frac{1}{\sqrt{N_xN_y}}\sum_{u=0}^{N_x-1}\sum_{v=0}^{N_y-1} F_{uv}e^{2\pi i \left(\frac{u x}{N_x} + \frac{v y}{N_y}\right)}  \label{fouriertrans2d5d}
    \end{align}

    where $x$ and $y$ represent the source coordinates and $u$ and $v$ represent the spatial frequencies. Using a Fast Fourier Transform~(FFT) algorithm allows calculation time in $O(N_xN_y\log{N_xN_y})$ where $N_x$ and $N_y$ are the $x$ and $y$ resolutions respectively~\cite{frigo2005design, carpenter2010graphics}.
    
\begin{figure}[tb]
        \centering
        {\includegraphics[trim={0 0 0 0},width=\linewidth,page=1]{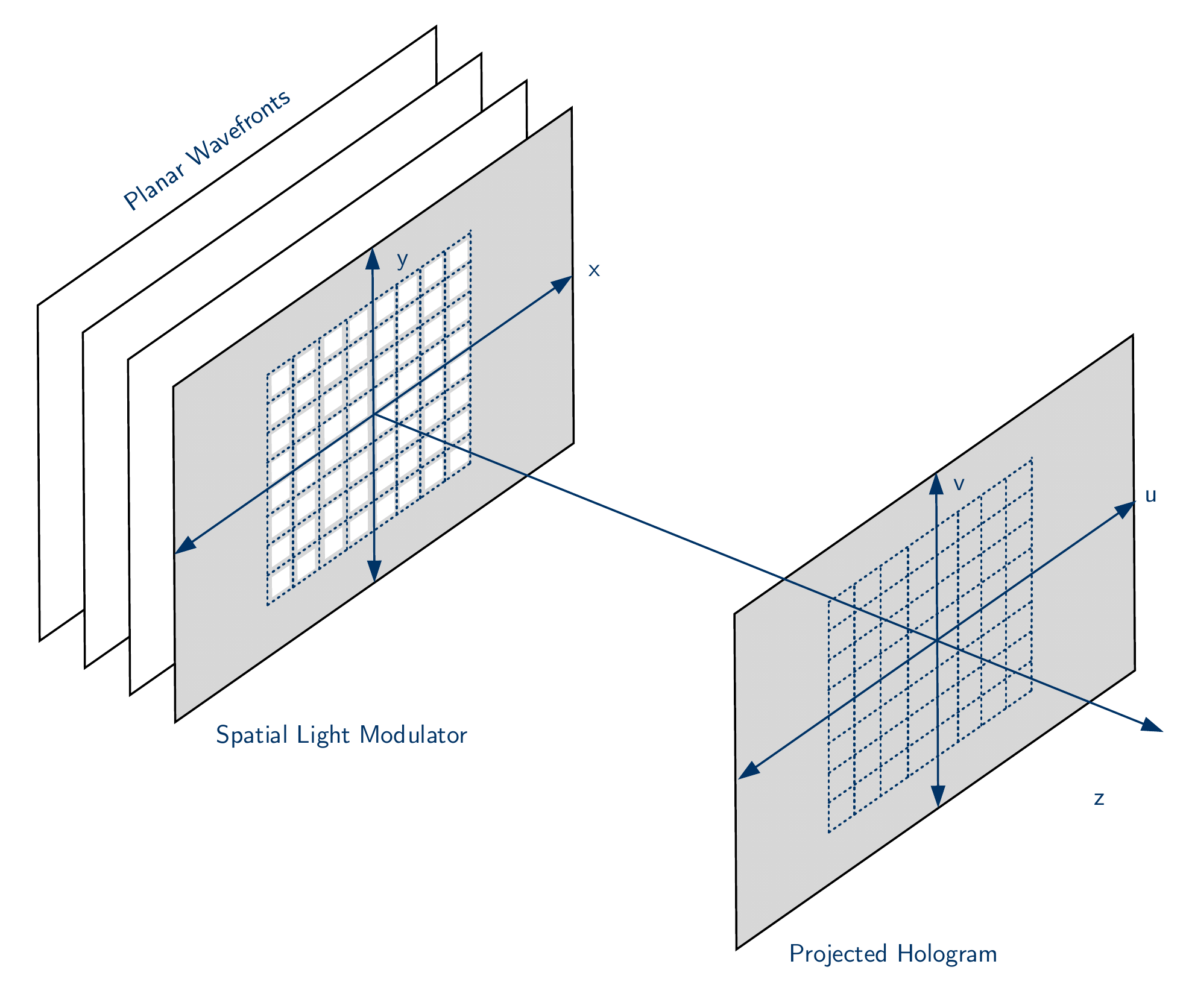}}
        \caption{Coordinate systems used in describing a hologram}
        \label{fig:HoloCoordinateSystems}
    \end{figure}

    It can be shown that when a coherent light source is passed through a Spatial Light Modulator~(SLM), the far-field pattern produced is equivalent to taking the DFT of the SLM aperture function multiplied by the incident illumination and a static pixel shape parameter~\cite{goodman2005introduction}. For a pixellated SLM acting on uniform intensity planar wavefronts  with $100\%$ fill factor pixels, the hologram produced is given by the DFT of that function as shown in Figure~\ref{fig:HoloCoordinateSystems}.
    
    In practical terms this means that finding an aperture function corresponding to a given far-field hologram is equivalent to finding a discrete function $f(x,y)$ where $F(u,v) = \mathcal{F}\{f(x,y)\}$. Mid-field or Fresnel holograms are similar in form with the addition of a quadratic phase rotation term.
    
    \begin{equation} 
    F_{u,v} = \underset{\scriptscriptstyle \text{Fresnel}}{\mathcal{F}}\{f_{x,y}\} = \underset{\scriptscriptstyle \text{Fraunhofer}}{\mathcal{F}}\{f_{x,y}e^{\frac{i \pi}{\lambda z}(x^2 + y^2)}\}
    \end{equation}
    
\begin{figure}[tb]
        \centering
        \begin{subfigure}[b]{0.33\textwidth}
            \includegraphics[width=\textwidth,page=25]{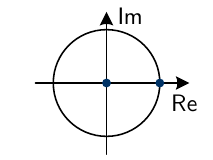}
            \caption{Binary Amplitude}
        \end{subfigure}
        \begin{subfigure}[b]{0.33\textwidth}
            \includegraphics[width=\textwidth,page=27]{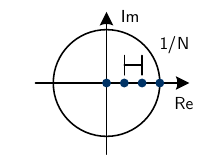}
            \caption{Multi-Level Amplitude}
        \end{subfigure}
        \begin{subfigure}[b]{0.33\textwidth}
            \includegraphics[width=\textwidth,page=29]{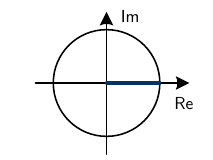}
            \caption{Continuous Amplitude}
        \end{subfigure}
        \begin{subfigure}[b]{0.33\textwidth}
            \includegraphics[width=\textwidth,page=26]{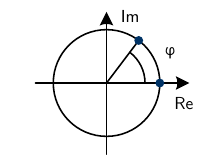}
            \caption{Binary Phase}
        \end{subfigure}
        \begin{subfigure}[b]{0.33\textwidth}
            \includegraphics[width=\textwidth,page=28]{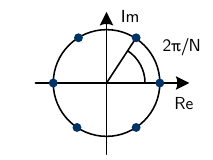}
            \caption{Multi-Level Phase}
        \end{subfigure}
        \begin{subfigure}[b]{0.33\textwidth}
            \includegraphics[width=\textwidth,page=30]{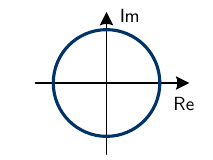}
            \caption{Continuous Phase}
        \end{subfigure}
        \caption[Common Modulation Schemes]{Common modulation schemes}
        \label{fig:modschemes}
    \end{figure}

    In real-world systems, SLMs are only capable of modulating the light in a limited manner, typically only in phase or amplitude \cite{Huang2018,deBougrenetdelaTocnaye:97}. Digital systems restrict this further to discrete energy levels. Some common SLM classes are shown in Figure~\ref{fig:modschemes}. 
    
    These restrictions have led to nearly as many algorithmic variants as there are implementations. One class of algorithm that are widely used when hologram quality is paramount are Holographic Search Algorithms~(HSAs) with DS and SA being the most common~\cite{DirectSearch_1,DirectSearch_2}.
    
    \section{Holographic Search Algorithms} \label{HSAs}
    
\begin{figure}[htbp]
        \centering
        {\includegraphics[trim={0 0 0 0},width=\linewidth,page=1]{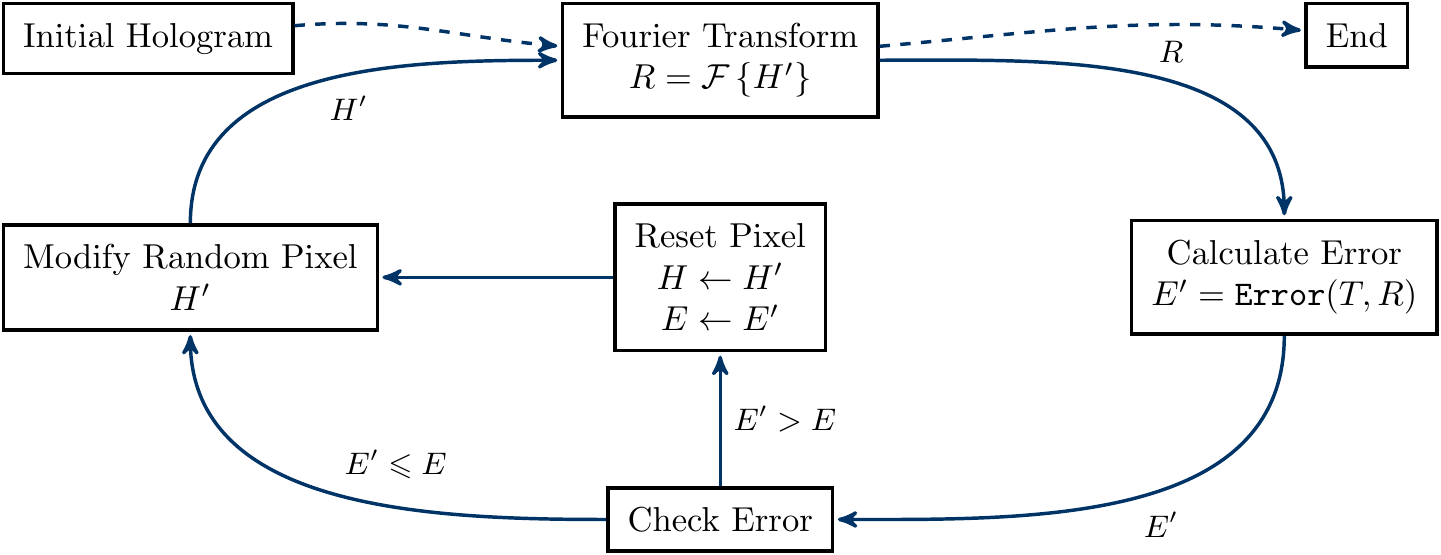}}
        \caption{Na\"ive Direct Search}
        \label{fig:dsslow}
    \end{figure}

    The DS approach is shown in Figure~\ref{fig:dsslow}. An initial guess at the hologram or SLM aperture function is taken. On each iteration of the algorithm a pixel is changed and a new replay field calculated. If this causes a reduction in the Mean Squared Error~(MSE) then the result is accepted, otherwise it is rejected and a new pixel modified~\cite{kirkpatrick1983optimization}. This approach is extremely slow but produces some of the best holograms when compared to alternative algorithms. For this work phase insensitive MSE is used for the $\texttt{Error}$ function where the error between the replay field $R$ and target image $T$ is given by:
    
    \begin{equation} \label{mse}
    \texttt{Error}(T,R) = \frac{1}{N_x N_y}\sum_{x=0}^{x=N_x-1}\sum_{y=0}^{y=N_y-1} \left[\abs{T(x,y)} -  \abs{R(x,y)}\right]^2.
    \end{equation}
    
\begin{figure}[htbp]
        \centering
        {\includegraphics[trim={0 0 0 0},width=\linewidth,page=1]{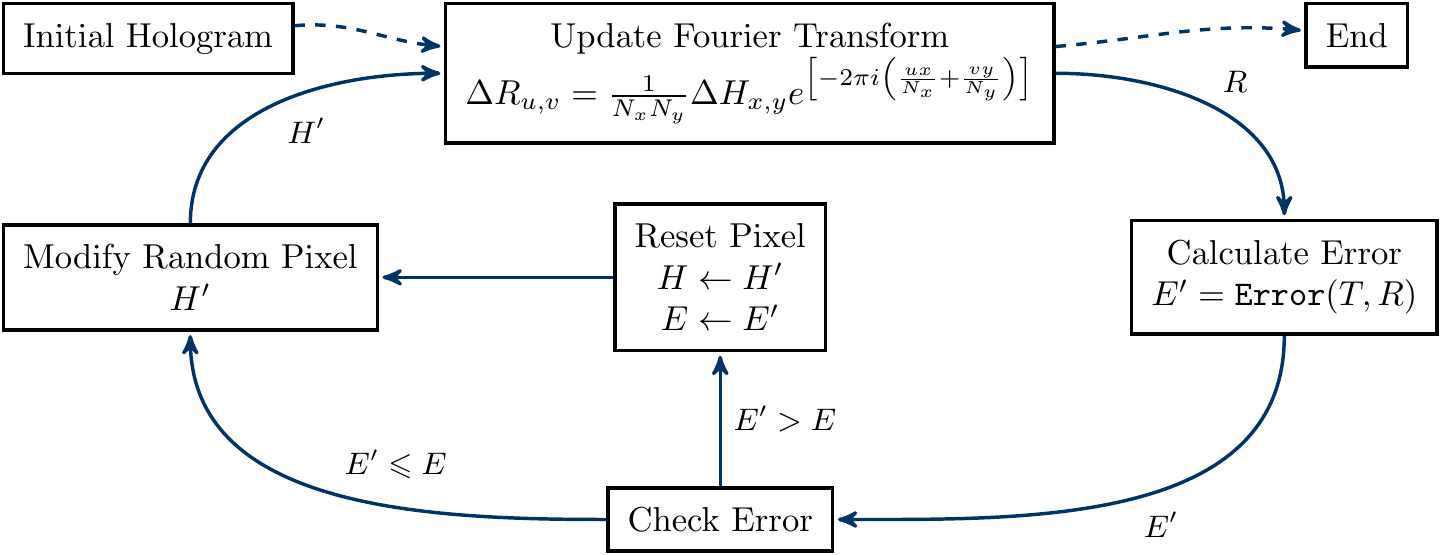}}
        \caption{Fast Direct Search}
        \label{fig:dsfast}
    \end{figure}

    A na\"ive approach to this algorithm would execute the 2D FFT on every iteration at a performance cost of $O(N_xN_y\log{N_xN_y})$. Knowing that system's initial state allows an update step that is merely  $O(N_xN_y)$ in cost, a significant saving,
    
    \begin{equation} \label{updatestep}
        \Delta R_{u,v} = \frac{1}{\sqrt{N_xN_y}}\Delta H_{x,y} e^{\left[-2\pi i\left(\frac{ux}{N_x}+\frac{vy}{N_y}\right)\right]}
    \end{equation}
    
    where $\Delta R_{u,v}$ is the change in replay field value caused by a change $\Delta H_{x,y}$ in aperture function. The updated algorithm is shown in Figure~\ref{fig:dsfast}.
    
\begin{figure}[htbp]
        \centering
        {\includegraphics[trim={0 0 0 0},width=\linewidth,page=1]{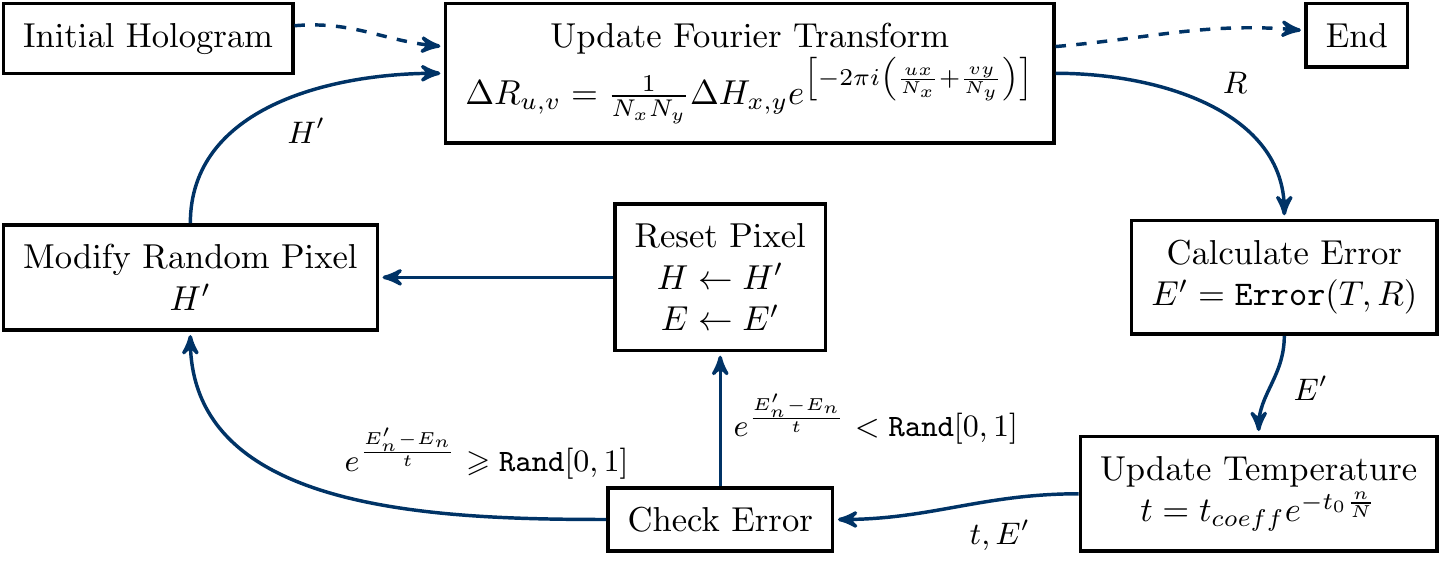}}
        \caption{Simulated Annealing}
        \label{fig:safig}
    \end{figure}

    In order to avoid the issue of converging to local minima found in greedy approaches such as DS, many researchers prefer the use of Simulated Annealing algorithms, Figure~\ref{fig:safig}, based on widely used simulated annealing techniques in computer science~\cite{Yang2009,Kirk1992}. This introduces an acceptance probability that will occasionally allow a pixel change that worsens the result and helps improve convergent image quality at the expense of increased convergence times~\cite{kirkpatrick1983optimization}. Most commonly the acceptance function is a modified Boltzmann function~\cite{dames1991efficient}.
    
    \begin{align} \label{boltzmann}
    P(\Delta E) & = e^{\frac{-\Delta E}{T}}        \\
    T           & = T_{\text{coeff}} e^{-T_0 \frac{n}{N}}
    \end{align}
    
    Both DS and SA approaches traditionally randomly select pixels, this work sets out to present a new paradigm that significantly improves convergence by using heuristic knowledge of the problem~\cite{carpenter2010graphics}.
    
    \section{Quantisation and Initial Guess}
    
    Search algorithm performance depends heavily on the initial guess for the diffraction field hologram. By far the dominant approach is to back-project the target using an inverse DFT. The limited modulation abilities of SLMs, Figure~\ref{fig:modschemes}, mean that the initial back projection requires a \textit{quantisation} step in order to adapt the complex valued function to the constraints of the display device. Each modification of a pixel value has an associated change in result, usually causing an increase in MSE.
    
    \begin{figure}[htbp]
        \centering
        {\includegraphics[trim={0 0 0 0},width=\linewidth,page=1]{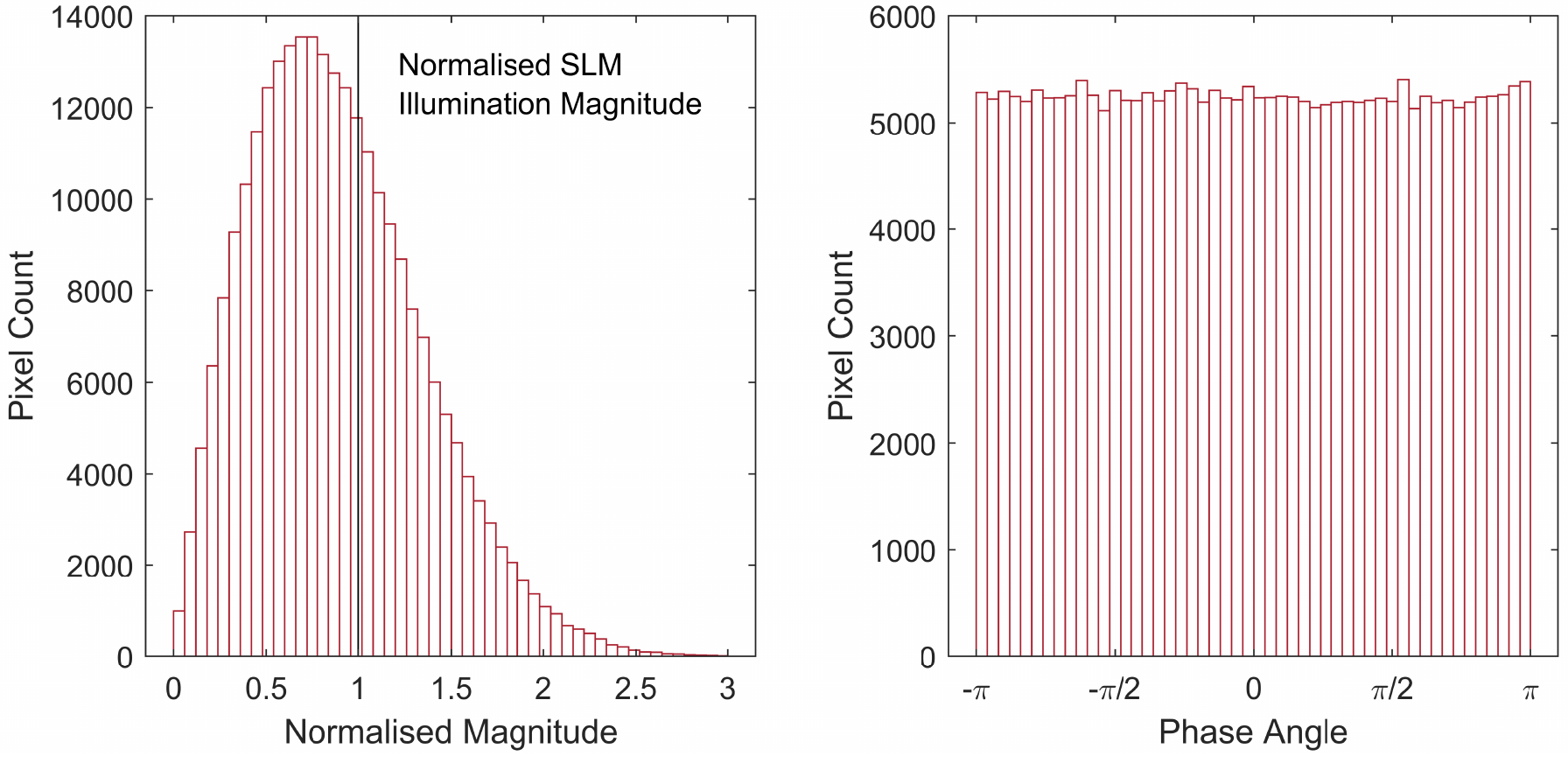}}
        \caption{Spread of magnitudes (left) and angles (right) of a back projected version of \textit{Mandrill}}
        \label{fig:hist1}
    \end{figure}
    
    \section{Heuristic Observations} \label{heuristic}
    
\begin{figure}[htbp]
        \centering
        {\includegraphics[trim={0 0 0 0},width=\linewidth,page=1]{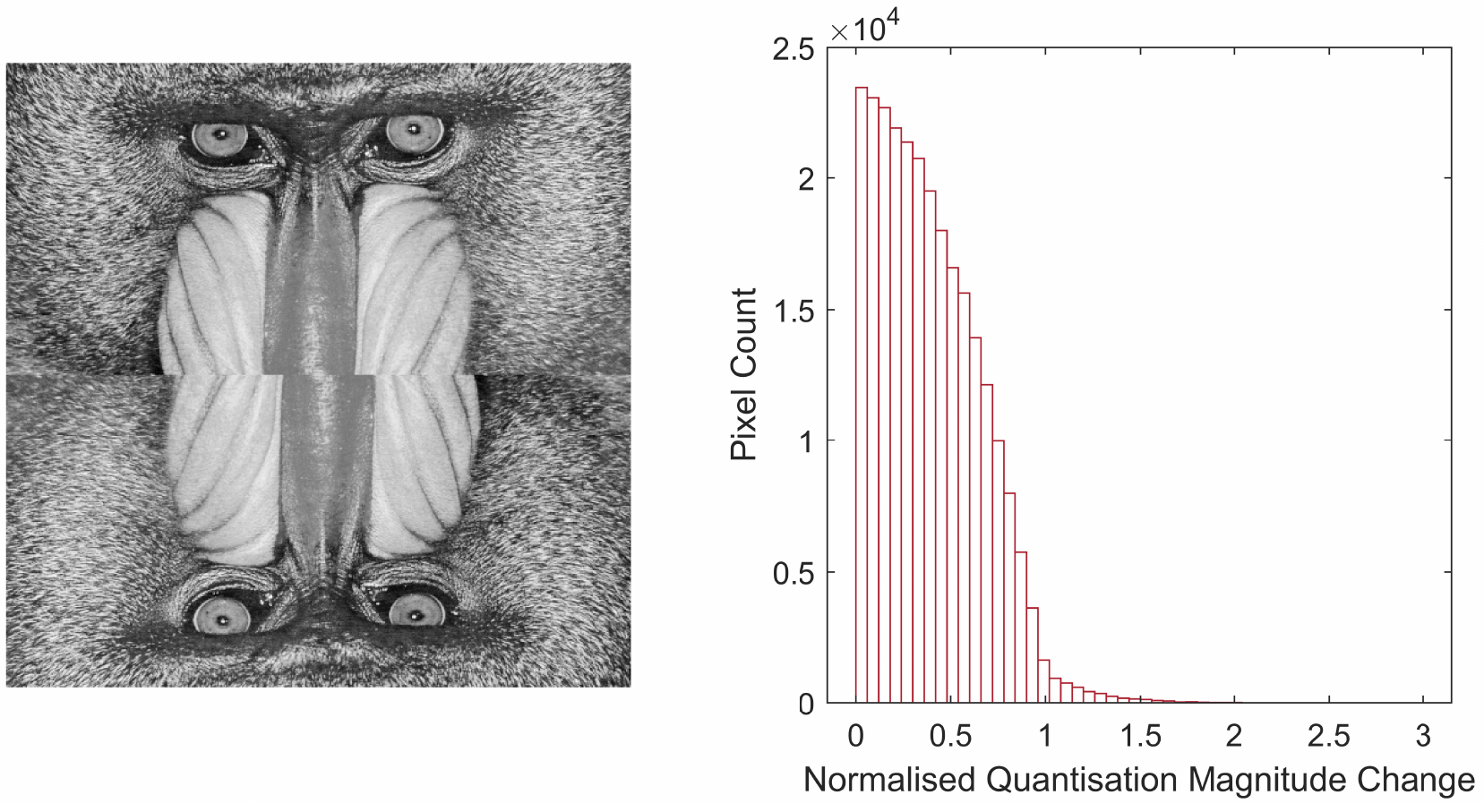}}
        \caption{Test image \textit{Mandrill} with artificially induced symmetry at 512 $\times$ 512 resolution (left) and spread of pixel magnitude changes due to quantisation for display on a pure phase device (right)}
        \label{fig:hist2}
    \end{figure}

    The back projection of a phase-randomised target image shows a spread in pixel magnitudes and phases, Figure~\ref{fig:hist1}. For the case of the standard \textit{Mandrill} test image, Figure~\ref{fig:hist2}(left), the normalised values are shown in Figure~\ref{fig:hist2}(right) where normalisation is taken as the equivalent of an illumination field of unit magnitude. The quantisation step will constrain these values to the SLM. In the case of phase holography, this will leave phase unchanged but will set all magnitudes equal to $1$ while in amplitude holography the amplitudes are set to the real values of the combined complex numbers. The artificially induced symmetry in the test image is to account for the rotational symmetry inherent in binary devices. While this is not ideal in terms of natural images, it is sufficient for our purposes.
    
\begin{figure}[htbp]
        \centering
        {\includegraphics[trim={0 0 0 0},width=\linewidth,page=1]{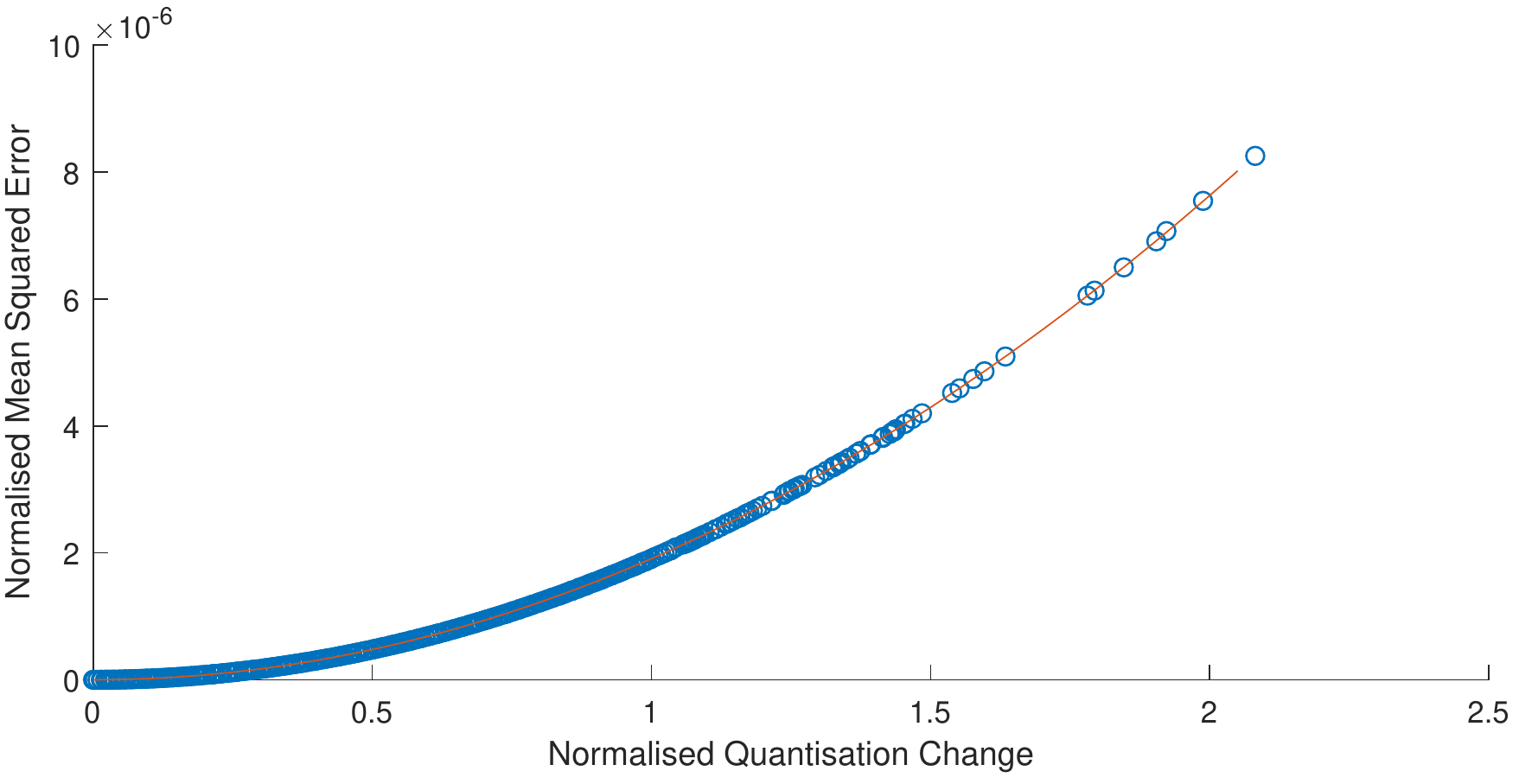}}
        \caption{Scatter plot of pixel value changes during quantisation of a continuous phase hologram against resultant error as well as expected square law relationship. Test image used is a 512 $\times$ 512 pixel version of Mandrill with induced rotational symmetry}
        \label{fig:hist3}
    \end{figure}

    The premise of this paper is based on the observation that the total change in error of the replay field caused by changing a single pixel is highly correlated to the magnitude of the SLM pixel value change during quantisation. Figure~\ref{fig:hist3} shows a scatter plot of the correspondence between independently quantising each of the $2.6\times10^5$ pixels of a continuous phase hologram and the resultant MSE. 
    
    Figure~\ref{fig:hist3} also shows the expected square law relationship due to the MSE error metric used. A correlation of $\gg0.99$ was observed between the trend line and the simulated dataset.
    
    Traditional search algorithms randomise the test pixel selection process. The relationship between change in pixel value during quantisation and the effect on the replay field suggests, however, that convergence will be improved by testing pixels with the greatest change during quantisation as they are heuristically likely to have greatest impact on error reduction.
    
    This approach is here termed Sorted Pixel Selection~(SPS) where, instead of randomly selecting test pixels, test pixels are chosen sequentially from a list in order of decreasing quantisation change.
    
    \section{Results} \label{results}
    
    \begin{figure}[htbp]
        \centering
        {\includegraphics[trim={0 0 0 0},width=\linewidth,page=1]{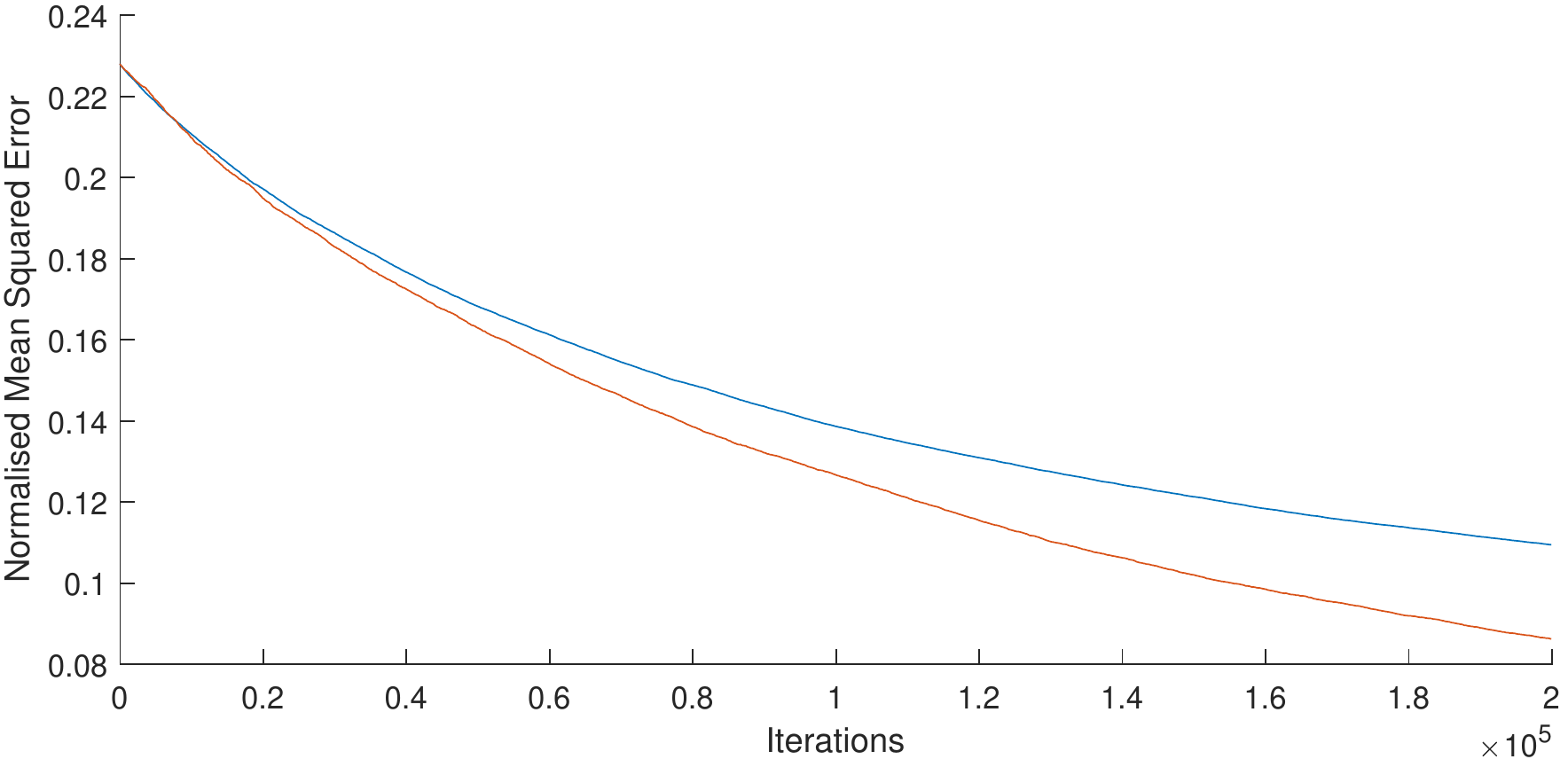}}
        \caption{Convergence of direct binary search with random pixel selection (blue) against sorted pixel selection (orange) for a $512 \times 512$ pixel \textit{Mandrill} test image being displayed on a binary phase spatial light modulator }
        \label{fig:hist4}
    \end{figure}
    
\begin{figure}[htbp]
        \centering
        {\includegraphics[trim={0 0 0 0},width=\linewidth,page=1]{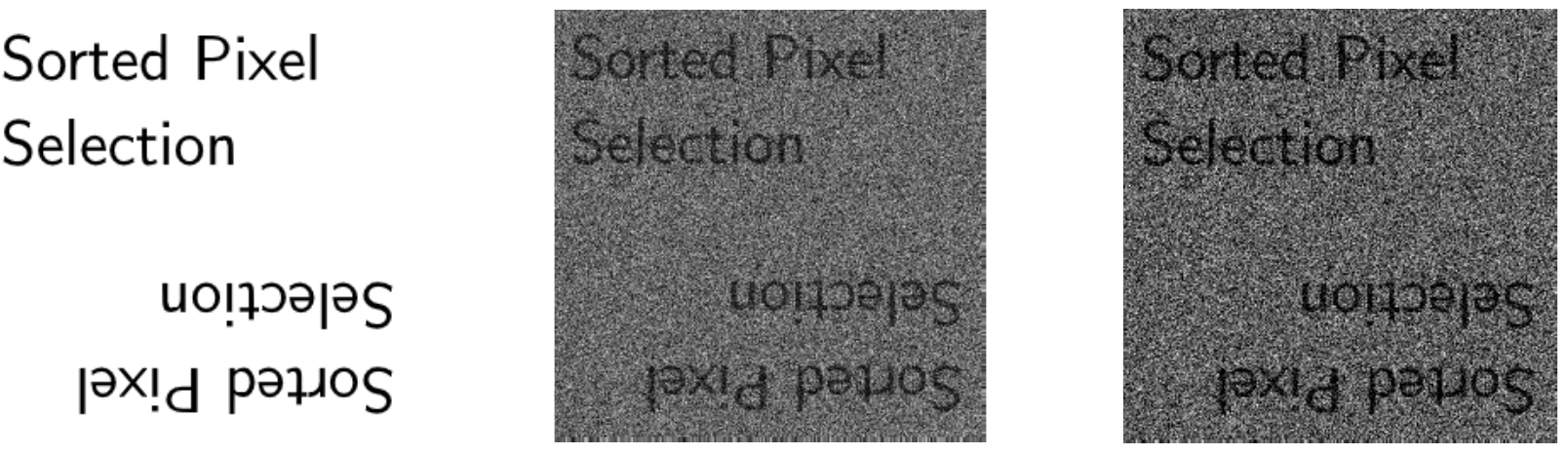}}
        \caption{Binary image generated with 10,000 iterations of direct search (centre) and direct search with sorted pixel selection (right) for a rotationally symmetric $256 \times 256$ pixel target image (left). There is a $19.1\%$ reduction in the mean squared error, when comparing the righthand image and the lefthand image.}
        \label{fig:hist5}
    \end{figure}

\begin{figure}[htbp]
        \centering
        {\includegraphics[trim={0 0 0 0},width=\linewidth,page=1]{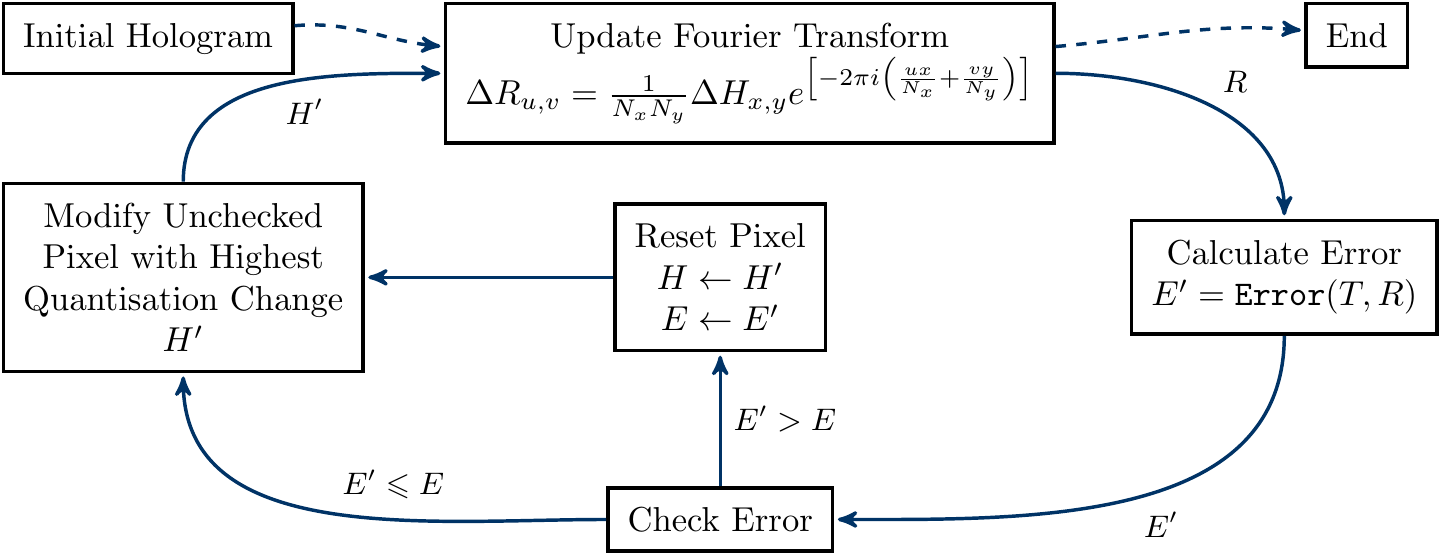}}
        \caption{Fast Direct Search with Sorted Pixel Selection}
        \label{fig:dsfastsorted}
    \end{figure}

    For example, modifying the direct search algorithm in Figure~\ref{fig:dsfast} to successively test pixels in order of decreasing quantisation change provides the convergence graph shown in Figure~\ref{fig:hist4} for a modified algorithm as shown in Figure~\ref{fig:dsfastsorted}. This uses the 512 pixel square \textit{Mandrill} test image on a binary phase SLM and exhibits a $16.5\%$ improvement in error reduction over $200,000$ iterations. Figure~\ref{fig:hist5} shows a comparison of the effect this has on a rotationally symmetric text image.
    
    This improvement is not massive but is significant due to the very low cost of implementation. There are many sorting algorithms that operate in $O(n\log{n})$ time where $n=N_xN_y$, which is the same complexity as the 2D FFT itself~\cite{knuth1997art}. Combined with their widespread availability makes adding a sorting step at the start of a search algorithm trivial. 
    
    Additionally, it is notable due to its stability. While Figure~\ref{fig:hist4} shows the case of showing \textit{Mandrill} on a binary phase SLM using $200,000$ iterations of a direct search algorithm, over $100$ tests showed improvements in the range of $[14.7\%,19.2\%]$ for a range of parameter combinations independent of:
    \begin{itemize}
        \item \textbf{Algorithm} - Simulated Annealing showed indistinguishable performance to the Direct Search case.  
        \item \textbf{Resolution} - Both algorithms experienced similar relative performance increases for image resolutions between 64 pixels square up to 2048 pixels square.   
        \item \textbf{Image} - While \textit{Mandrill} was the primary test image used, tests with other images from the USC-SIPI image database~\cite{sipidatabaseref}. This included tests with the USAF target with a very different initial power spectral density.
        \item \textbf{Iterations} - The relative performance gain was consistent at $\sim 15\%$ regardless of the number of iteration count.   
        \item \textbf{Modulation Scheme} - While the graph shown in Figure~\ref{fig:hist4} is for the case of a pure phase modulator, no difference in relative performance improvement was observed for the case of a pure amplitude modulator. The case of hybrid phase-amplitude modulation schemes has not yet been rigorously tested.
        \item \textbf{Modulation Levels} - Search algorithms are primarily used in the case of binary holograms as competing algorithms such as Gerchberg-Saxton offer better performance in the multi-level or continuous cases. Running tests for 8, 256 and continuous modulation levels showed no significant change in relative convergence improvement over the binary case.
    \end{itemize}

    The technique discussed here has been applied to far-field or Fraunhofer holograms. It is expected to be equally applicable to mid-field or Fresnel holograms as the addition of a phase rotation term does not change the heuristic arguments made earlier.

    One final observation of note is that while error reduction showed a $\sim 15\%$ improvement, the total number of pixel changes accepted remained largely the same with the impact of successful pixel modifications rising in the SPS case.

    \section{Comparison with Previous Work} \label{comparison}

    In this paper we presented a novel method of using the quantisation change to improve the family of search algorithms used in CGH. 
    
    Quantisation has been a focus of improvements in holography since the early 70's and it has been recognised that quantisation changes cause a noise term in the reconstruction for several decades~\cite{Dallas:74}. 
    
    A number of papers have examined the expected proportion of error that will be caused by quantisation as opposed to other limiting factors such as energy conservation and a number of non-iterative algorithms have been developed for reducing this quantisation noise~\cite{Rines:81}. Non-iterative approaches have included adjusting point source locations spatially in the diffraction field~\cite{Hsueh:78} and the use of dummy areas with variable regions of interest~\cite{Akahori:86}. 
    
    Common iterative algorithms such as Gerchberg-Saxton~(GS) deal with quantisation error as part of the expected whole~\cite{FienupXX}. Some sources have discussed quantisation error independently of other sources. For example, Yang et. al. discussed quantisation error in off-axis configurations~\cite{Yang:00}. HSAs are typically applied in applications with binary or low quantisation levels and quantisation is known to be a major error source~\cite{kang}. A number of alternative quantisation techniques such as error diffusion have been used to attempt to reduce this as well as replay field manipulation approaches.\cite{Yang:00, sdfsdf}
    
    To our knowledge no previous work has directly linked the quantisation change with expected error on a pixel-by-pixel basis. Figure~\ref{fig:hist3} allows us to suggest, for the first time, a modification to popular search algorithms that improves performance with limited additional overhead.

    \section{Limitations and Future Investigation}
    
    This paper has considered only a small subset of holographic search algorithms. While the proposed approach is postulated to be equally efficacious in other cases, this has yet to be tested. This technique has also only been applied to phase insensitive, binary phase Fraunhofer holograms. Further study is required to examine its applicability to phase sensitive applications, alternative devices and the Fresnel region.
    
     While this paper has sorted purely by quantisation change, it is likely that other variables will also have significant impact. It is expected that other features such as position in the replay field are likely to make a significant impact.
    
    Also necessary of investigation is the behaviour when the number of iterations is higher than the number of pixels. Both of these issues can be addressed by use of the development of a probabilistic function that weights a number of parameters including quantisation change during the pixel selection step. Instead of iterating through a sorted list, the function would probabilistically select pixels with a greater chance of impact.
        
    \section{Conclusion}
    
    This work has presented a modification to existing holographic search algorithms with a relative convergence error reduction improvement in the range of $[14.7\%,19.2\%]$. Tests were run for direct search and simulated annealing algorithms as well as a range of test images, parameters and spatial light modulators with very similar performance improvements in all cases.
    
    Unlike traditional search algorithms where test pixels are selected randomly, this paper has presented a sorted variation where the pixels are sorted by the magnitude of the quantisation change immediately after back-projection. This was initially justified by some heuristic observations of the average nature of the back projected image and then trialled for a range of tests set-ups. While this is a general modification applicable to many HSAs and should not be regarded as a separate algorithm, it is proposed that this technique be referred to as Sorted Pixel Selection~(SPS).
        
    While the performance improvement is small, it is consistent across a wide variety of test cases as well as being cheap and easy to apply to existing set-ups. For large images, search algorithms can take many hours to run and the improvements observed offer a significant benefit.
        
    \bibliography{references}
    
\end{document}